\documentclass[12pt,onecolumn,draftcls]{IEEEtran}

\ifCLASSINFOpdf
\else
\fi
\usepackage{graphicx}
\usepackage{epstopdf}
\usepackage{stfloats}
\usepackage{bm}
\usepackage[cmex10]{amsmath}
\usepackage{algorithmic}
\usepackage{algorithm}
\usepackage{array}

\usepackage{mdwmath}
\usepackage{mdwtab}
\usepackage{multirow}

\usepackage{amsfonts}

\usepackage{amssymb}

\usepackage{color}

\begin{document}
%
\title{Fundamentals of Physical Layer Anonymous Communications: Sender Detection and Anonymous Precoding}

\author{Zhongxiang Wei,~\IEEEmembership{Member,~IEEE,}
        Fan Liu,~\IEEEmembership{Member,~IEEE,}
            Christos Masouros,~\IEEEmembership{Senior Member,~IEEE,}
            and H. Vincent Poor,~\IEEEmembership{Life Fellow,~IEEE}
\thanks{Zhongxiang Wei, Fan Liu, and Christos Masouros are with the Department of Electronic and Electrical Engineering at the University College London, London, UK. Email: \{zhongxiang.wei, fan.liu, c.masouros\}@ucl.ac.uk}
\thanks{H. Vincent Poor is with the Department of Electrical Engineering, Princeton
University, Princeton, NJ 08544 USA (e-mail: poor@princeton.edu).}
\thanks{This work was supported by the Engineering and Physical Sciences Research Council, UK, under project EP/R007934/1.} 

}

\maketitle

\begin{abstract}

In the era of  big data, anonymity is recognized as an important attribute in privacy-preserving communications. The existing anonymous authentication and routing are applied at higher layers of  networks, ignoring  physical layer (PHY) also contains privacy-critical information. 
In this paper, we introduce the concept of PHY anonymity, and reveal that the receiver can unmask the sender's identity by only analyzing the PHY information, i.e., the signaling patterns  and the characteristics of channel.
We investigate two practical scenarios, where the receiver has more antennas than the sender in the strong receiver case, and vice versa in the strong sender case. For each scenario, we first investigate sender detection strategy at the receiver, and then we develop corresponding anonymous precoding to address anonymity while guaranteeing high receive signal-to-interference-plus-noise-ratio (SINR) for communications.
In particular, an interference suppression anonymous  precoder is first proposed, assisted by a dedicated transmit phase equalizer for removing phase ambiguity. Afterwards, a constructive interference anonymous precoder is further investigated to utilize the inter-antenna interference as a beneficial element without loss of sender's anonymity.
Simulation demonstrates the proposed anonymous precoders are able to preserve the sender's anonymity and simultaneously guarantee high receive SINR, opening a new dimension on PHY anonymous designs.

\end{abstract}


%
\IEEEpeerreviewmaketitle

\section{Introduction}

In the era of cloud computing, storage and communications, the  misuse of confidential data, such as e-health analytics and infrastructure monitoring, has attracted much attention in both commercial and military applications. Due to the inherent broadcast nature of wireless communications,  threats arise from two main aspects, namely security and privacy.
The aim of security is to prevent the confidential signal from being eavesdropped upon by potential adversaries. There has been extensive research on cryptography \cite{Massey1988An},  authentication \cite{Zhang2016Efficient}, covert communication \cite{Hu2018Covert}, multiple-input and multiple-output (MIMO)  beamforming plus artificial noise design \cite{Wei2020Device}, and cooperative jamming \cite{Dong2010Imporving}, from the upper layer to the physical layer (PHY) of networks. These extensive works enable confidential communications among the legitimate parities, while ensuring the signal is not breakable and decodable at adversaries \cite{Herfeh2020Physical}.
In contrast, the aim of privacy is to guarantee the communication quality of the legitimate users, and meanwhile to conceal the identities of communication parties or the specific users' participation during the communications, also defined as anonymous communications \cite{Serjantov2002Towards}.
A typical example of anonymous communications arises from the upcoming cloud computing systems, where users offload their data to the edge, i.e.,  cloud servers at access points (AP)s, to avoid heavy computing workloads and prolong their battery lifetime. Nevertheless, as these data may be  personal and confidential, it is expected that APs only process the offloaded data without extracting the sender's identity.  
Another example comes from remote healthcare applications, where patients wish to anonymously access online medical services by only sharing bio-information, whereas all remaining private information and the user's identity must be kept unknown. 
In summary, privacy-preserving techniques have become imperative, where the communication parties should only be able to process data without the knowledge of other participants' identities.  


There are three categories of anonymity, namely sender anonymity, receiver anonymity and bi-directional anonymity. Sender anonymity means  the receiver cannot trace the
sender’s identity; receiver anonymity means  the sender can contact the receiver without knowing its identity; while bi-directional anonymity means  both the sender and receiver communicate without knowing each other’s identities \cite{Danezis2006Introducing}.
To this end, researchers have unveiled various ways to enhanced anonymity at high layers of networks, such as authentication and routing protocols. 
Anonymous authentication schemes have been proposed for cellular networks \cite{Chang2009} \cite{Youn2009Weakness}, wireless body area networks \cite{Liu2014Certificateless},  wireless local area networks \cite{Lu2009A}, device-to-device (D2D) communications \cite{Gope2019Anonymous}, Radio Frequency IDentification \cite{Ren2013An}, and other authentication protocol designs \cite{Emura2016Secure} \cite{Lian2015Periodic}. 
The general design principle is to apply anonymous  authentication \cite{Chang2009} \cite{Liu2014Certificateless}, mutual authentication \cite{Lu2009A}, distributed authentication \cite{Gope2019Anonymous}, or multi-round authentication/encryption \cite{Lian2015Periodic} to conceal the participants' identities. 
In \cite{Chang2009}, the authors proposed anonymous authentication schemes for roaming service in global mobility networks, and subsequently Youn  \textit{et al.,} \cite{Youn2009Weakness}  proposed different attack schemes, by which an entity can compute all session keys of a mobile user to recover the users' identities. 
In \cite{Liu2014Certificateless}, an anonymous authentication scheme was proposed for wireless body area networks, based on a novel certificate-less signature scheme. 
In \cite{Lu2009A}, a link layer mutual authentication scheme was proposed to enhance the users' location privacy in wireless local area networks. 
In \cite{Gope2019Anonymous}, a distributed authentication protocol was proposed for D2D fog computing, where devices proximate to others offer help to obtain faster authentication without involving any centralized server.
The authors in \cite{Emura2016Secure} proposed an identity-based encryption technique for encrypting packets, where  users can conceal their IP addresses from service providers. 
In \cite{Lian2015Periodic},  a periodic $K$-times anonymous authentication system was investigated, where a user can anonymously show credentials at most $K$ times in one time period. 
On the other hand, a great deal of effort has been invested in designing anonymous routing for the Internet and ad
hoc networks \cite{Basagni2001Secure} \cite{Sakai2019On}. The fundamental principle is to  preserve the privacy of end hosts as well as routing paths  by a number of encrypted layers, where routers serve as proxies and any given intermediate nodes are unaware of where the source and sink of the message are located. 

Nevertheless, there are still issues left to be addressed by the aforementioned anonymizing techniques.
i) Since the existing anonymous, distributed, or multiple times authentication techniques are generally based on  public-key cryptosystems \cite{Chang2009}, they may be  restrictive in many emerging scenarios of 5G-beyond networks due to the high computational requirement and latency, where  cooperation among the entities is required for both ring and group signatures \cite{Fang2019Machine}.
Although  smaller key sizes have been proposed based on elliptic curve cryptosystems, the users still need extra computation to verify the certificates of others, and a pool of certificates is required for the certification authority for maintaining users’ keys. More importantly, distributed and multi-round authentication/encryption techniques require the cooperative participants to be fully trusted and honest. 
ii) The existing anonymous routing protocols are only applicable for large-scale networks, where multiple cooperative users are involved to guarantee anonymity.
Hence, none of the users could be off-line during the underlying process, and it is vulnerable to the internal malicious member attacks that can easily break the anonymity.
iii)  The premise of authentication/encryption schemes relies on the condition that the adversaries have limited processing capability, which is constantly being surmounted due to the growth of computational power and quantum computing. 
iv) The existing anomymizing techniques and associated protocols are  employed at the upper layers of networks, assuming PHY provides a privacy-preserving link.  
In fact, the PHY also contains
information that can be used to extract the nodes' identities. When an anonymously authenticated/encrypted sender transmits a signal via its wireless channel, the recipient can analyze the signaling patterns based on the characteristics of channel fading, and then is able to unmask the origin of the received signal at the PHY directly. 
Thus, privacy threats start from the acquisition of data, which necessitates complementary privacy techniques that reside at the PHY. 
Note that while the existing PHY location verification and identity authentication techniques \cite{Fang2019Machine} \cite{Brighente2019Machine} leverage the physical properties of the wireless medium as a  source of domain-specific information to complement security mechanisms, they aim at preventing legitimate transceivers from being spoofed/attacked by external eavesdroppers.
Since they do not provide anonymity for legitimate communication parties, they are encapsulated in the set of traditional PHY security rather than anonymity solutions \cite{Mukherjee2014Principles}.

Motivated by the aforementioned open challenges, in this paper, we present a first attempt to exploit PHY sender detection schemes and their counterpart anonymous precoding techniques. Our contributions are summarized as follows.

\begin{enumerate}[]
	\item  This is the first work to unveil that PHY information, i.e., the signaling pattern and the inherent characteristics of channel fading, can be judiciously analyzed to unmask sender's identity and this incurs an unprecedented vulnerability  by anonymity-violating behavior at the receiver. Focusing on different antenna configurations, we further propose two novel sender detection strategies that only exploit  the PHY information to break the sender's anonymity. For the strong receiver case where the number of receive antennas is larger than the transmit antenna of the sender, a maximum likelihood estimation (MLE) based sender detector is proposed. While in the strong sender case with the reduced receive degrees-of-freedom (DoF) in detection, we further propose a maximum norm (M-Norm) based  sender detector, with lower computational complexity over the MLE based detector.

	\item For both antenna configurations, a series of corresponding anonymous PHY precoding techniques is proposed against the sender detection schemes.  We first propose an interference suppression based anonymous (ISA) precoder that maximizes per-antenna SINR performance while simultaneously addressing the sender's anonymity, assisted by a dedicated transmit phase equalization design for eliminating phase ambiguity. We also prove that the applied semi-definite relaxation (SDR) in optimization is tight and the optimality of the  precoder is always maintained.
	
	\item Then, we further propose a constructive-interference (CI) based anonymous (CIA) precoder, which is able to utilize inter-antenna interference as a beneficial element for further enhancing receive quality without loss of the sender's anonymity. Importantly, the CI based anonymous precoder enables multiplexing more data streams than the number of transmit antennas, and hence is also applicable to  
	the strong receiver case. 
	\end{enumerate}

Our study also reveals a number inherent properties of the anonymous precoding designs.
\begin{enumerate}[]
     
	\item Sender anonymity is achieved at the cost of reduced receive diversity, and hence the conventional receive equalizer that relies on the  deterministic channel information becomes inapplicable.
	\item  Receive phase equalization, typically employed in classical optimization-based precoders, is not achievable in anonymous communications and thus transmit phase equalization is essential for correct demodulation at the receiver.
	\item The two aims of optimizing quality-of-service (QoS) and preserving PHY layer anonymity are conflicting and therefore there exists a non-trivial trade-off between improving receive quality and guaranteeing sender's anonymity.
\end{enumerate}

\textit{Notation}:
Matrices and vectors are represented by boldface capital and lower case letters, respectively. $\vert\cdot \vert$ denotes the absolute value of a complex number. $\vert\vert\cdot \vert\vert$ denotes the Euclidean  norm. $\bm{A}^T$, $\bm{A}^H$ and Tr$(\bm{A})$ denote the transpose, Hermitian transpose and trace of a matrix $\bm{A}$. Rank($\bm{A}$) denotes the rank of a matrix $\bm{A}$. diag ($\bm{A}$) returns a diagonal matrix with diagonal elements from a matrix $\bm{A}$. $\bm{A}\succeq 0$ means $\bm{A}$ is a positive semi-definite matrix. $\Re$ and $\Im$ denote the real and imaginary parts of a complex variable. $\bm{I}_n$ means an $n$-by-$n$ identity matrix.

\section{System Model and Anonymity Performance Metrics}
In this section, the system model and anonymity performance metric are presented in subsections II-A and II-B, respectively.

\begin{figure}
	\centering
	\includegraphics[width=3.1 in]{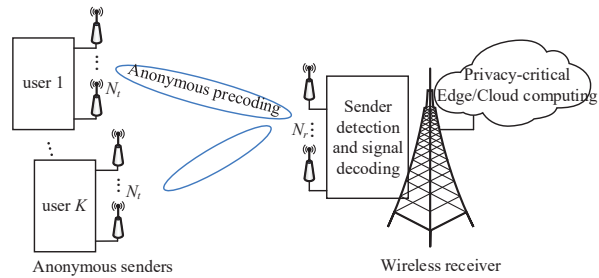}
    \caption{Illustration of  system model, where $K$ users transmit data to the receiver with QoS and sender's anonymity requirements.}
    \label{fig:systemmodel} 
\end{figure}

\subsection{System Model}

We consider an uplink multiuser MIMO system depicted in Fig. 1, and in particular a sender anonymity scenario, where users anonymously transmit data to an AP receiver without leaking their identities.
Assume the user set $\mathbb{K}$ consists of $K$ users ($|\mathbb{K}|=K$), and there is one user communicating (denote $\mathbb{S}$ as the sender) with the receiver at each time slot in a time-division-multiple-access (TDMA) fashion. 
The receiver is equipped with $N_r$ receive antennas, while each user is equipped with $N_t$ transmit antennas. 
Define $\bm{H}_k\in \mathbb{C}^{N_r \times N_t}$ as the MIMO channel between the user $k$ and receiver, $\forall k \in K$.
Define $\bm{W}_k$ as the precoding matrix and $\bm{s}_k$ as the symbol vector to be transmitted by the $k$-th user. The received signal at the receiver is written as 


\begin{small}
\begin{equation}
\begin{split}
\bm{y}=\bm{H}_k \bm{W}_k \bm{s}_k+\bm{n},
\label{eq:Recevied signal 1}
\end{split}
\end{equation}
\end{small}%
where $\bm{n}\in \mathbb{C}^{N_r \times 1}  $ denotes the circularly symmetric complex Gaussian (CSCG) noise at the receiver, and its $r$-th element follows $[\bm{n}]_r \sim \mathcal{CN}(0,\sigma^2)$, $\forall r \in N_r$.

\subsection{Performance Metric of Anonymity}

Higher layer anonymity is typically  quantified by an entropy based metric \cite{Chou2007An}.
Considering the set $\mathbb{K}$,  the receiver decides/assigns each user $k$ in $\mathbb{K}$ a probability $p_k$ of being the sender $\mathbb{S}$ based on its sender detection strategy. Hence, the anonymity entropy can be calculated as 

\begin{small}
\begin{equation}
\begin{split}
H(\mathbb{K})=-\sum_{k\in \mathbb{K} }^{}p_k\mathrm{log}_2p_k.
\label{eq:HK}
\end{split}
\end{equation}
\end{small}%

Evidently, the maximum anonymity entropy $H_{\mathrm{max}}(\mathbb{K})=\mathrm{log}_2(K)$ is achieved when $p_k = \frac{1}{K}, \forall k \in \mathbb{K}$, 
i.e., the users are equally likely senders. Hence,  the sender detection and sender's anonymity-preserving strategies are motivated by the following Remark 2.1.

\textbf{\textit{Remark 2.1}}:
As suggested by \eqref{eq:HK}, the sender detection  (denote as $\mathcal{D}$) for the receiver is to correctly identify the real sender $k$ with a high probability $p_k$ of being the sender, i.e.,
\begin{small}
\begin{equation}
\begin{split}
\mathcal{D}^{\ast}=\mathrm{max}(p_k|\mathbb{S}:k),
\label{detection strategy}
\end{split}
\end{equation}
\end{small}
On the other hand, a favorable sender's anonymity-preserving design at the user side is to deteriorate the  sender detection performance at the receiver, while guaranteeing reasonable receive quality for communication signal. $~~~~~~~~~~~~~~~~~~~~~~~~~~~~~~~~~~~~~~~~~~~~~~~~~~~~~~~~~~~~~~\square$


In the following, we will first reveal the sender detection design at the receiver in Section III, and accordingly the anonymous design at the sender is designed in Section IV, respectively.

\section{Sender Detection Strategy }
In this section, we study the sender detection schemes at the receiver. 
Since the receiver only analyzes the PHY information, i.e., the received signal and the inherent characteristics of the wireless channels  to disclose identity of the sender, under the TDMA premise the sender detection can be formulated as a multiple hypothesis testing (MHT) problem

\begin{small}
\begin{equation}
\begin{split}
\bm{Y}=\left\{
\begin{array}{rcl}
\mathcal{H}_0:   &  \bm{n}, \\
\mathcal{H}_1:   &  \bm{H}_1\bm{W}_1 \bm{s}_1+\bm{n},\\
\vdots\\
\mathcal{H}_K:   &    \bm{H}_K\bm{W}_K \bm{s}_K+\bm{n},
\end{array} \right.
\label{eq:HT0k}
\end{split}
\end{equation}
\end{small}%
where the hypothesis $\mathcal{H}_0$ means no data is transmitted from the user set $\mathbb{K}$ and only noise appears at the receiver. In comparison, hypothesis $\mathcal{H}_k$ means there is a signal coming from the $k$-th sender. 
Hence, the receiver attempts to detect the correct hypothesis from the $1+K$ MHT candidates.
Apparently, to handle the MHT problem, the receiver can first detect whether the hypothesis $\mathcal{H}_0$ is true or false, and  only turns to detect the origin of the signal (the hypotheses $\mathcal{H}_1$ to $\mathcal{H}_K$) when  $\mathcal{H}_0$ is decided as a false hypothesis. 

The detection of $\mathcal{H}_0$ leads  to the classic energy detection that has been extensively researched in the context of cognitive radios \cite{Liang2008Sensing}  \cite{Zeng2009Eigenvalue}, which is  briefly discussed for the sake of completeness. Based on the received signal $\bm{y}$, the test statistic for the energy detector is given by

\begin{small}
\begin{equation}
\begin{split}
\mathcal{T}(\bm{y}) = \frac{1}{N_r} \sum_{n=1}^{N_r} ||\bm{y}(n)||^2= \frac{||\bm{y}||^2}{N_r},
\label{eq:Ty}
\end{split}
\end{equation}
\end{small}%

Under hypothesis $\mathcal{H}_0$, the test statistic $\mathcal{T}(\bm{y})$ is a random variable whose probability density function (pdf) is a Chi-square distribution with $2N_r$ DoF. 
Define the  probability of false alarm as,
under hypothesis $\mathcal{H}_0$, the probability of the receiver falsely declaring the presence of an incoming signal.
Assume a detection threshold $\beta$, and the probability of the false alarm of $\mathcal{H}_0$ is then approximated by

\begin{small}
\begin{equation}
\begin{split}
P_f(\beta|\mathcal{H}_0 ) = \mathrm{ Pr} (\mathcal{T}(\bm{y}) > \beta|\mathcal{H}_0) = \int_{\beta}^{\infty}  \psi_{(2N_r)} (x)~ \mathrm{d} x,
\label{eq:Pf}
\end{split}
\end{equation}
\end{small}%
where $\psi_{(2N_r)}(x)$ denotes the pdf of a Chi-square
distributed variable with $2N_r$ DoF.
Note that there is a multitude of advanced energy detection schemes, such as eigenvalue-based detection \cite{Zeng2009Eigenvalue} and feature detection \cite{Kosunen2013Survey}. Since energy detection has been extensively researched and is not our main contribution, we refer readers to \cite{Zeng2009Eigenvalue} \cite{Kosunen2013Survey} for details.
Once the receiver has sensed the presence of an incoming signal, it turns to detect the origin of the received signal, and we have the following Remark 3.1 for the sender detector design at the PHY.  

\textbf{\textit{Remark 3.1}}:
The detection of the user's identity in the TDMA scenario is equivalent to the identification of the propagation channel (which is also the unique and unchangeable PHY identity of the  user)  from the received signal.
Hence,  the receiver is able to utilize the characteristics of the MIMO channel  to disclose the sender. ~~~~~~~~~~~~~~~~~~~~~~~~~~~~~~~~~~~~~~~~~~~~~~~~~~~~~~~~$\square$

Since the characteristics of the MIMO channel (i.e., the dimension and transmit/receive diversity) depend on the configurations of  $N_r$ and $N_t$, in the following we consider the strong receiver  ($N_r > N_t$) and strong sender  ($N_r \leq N_t$) cases and design specific detector for each.

\subsection{The Case of a Strong Receiver ($N_r > N_t$)}

This configuration is a common scenario at uplink transmission since an AP or base station is normally equipped with more antennas than a user.
Recalling the MHT in \eqref{eq:HT0k}, since the receiver has sensed the received signal $\bm{y}$ and has the knowledge of CSI set $\bm{H}_k, \forall k\in \mathbb{K}$, it is easy for the receiver to apply the maximum likelihood estimation (MLE) to disclose the estimate of the  transmitted vector $\bm{x}_k=\bm{W}_k\bm{s}_k\in \mathbb{C}^{N_t \times 1}$ as
\begin{small}
\begin{equation}
\begin{split}
\hat{\bm{x}}_k=\bm{H}_k^{\dag}\bm{y}= \bm{W}_k \bm{s}_k+\bm{H}_k^{\dag}\bm{n},
\label{eq:hatxk}
\end{split}
\end{equation}
\end{small}%
where $\bm{H}_k^{\dag}=(\bm{H}_k^H\bm{H}_k)^{-1}\bm{H}_k^H$ denotes the pseudo-inverse of the channel $\bm{H}_k$.
Then, the estimated vector $\hat{\bm{x}}_k$ is  multiplied by $\bm{H}_k$ to imitate that it propagates through $\bm{H}_k$, and 
a re-constructed signal $\hat{\bm{y}}_k$ is obtained as


\begin{small}
\begin{equation}
\begin{split}
\hat{\bm{y}}_k=\bm{H}_k \hat{\bm{x}}_k=\bm{H}_k\bm{W}_k \bm{s}_k+\bm{H}_k\bm{H}_k^{\dag}\bm{n}.
\label{eq:hatyk}
\end{split}
\end{equation}
\end{small}%

Apparently, if the received signal indeed comes from the $k$-th user (which propagates from the channel $\bm{H}_k$), the re-constructed signal $\hat{\bm{y}}_k$ built on  $\bm{{H}}_k$ should have the smallest Euclidean distance to the actual received signal $\bm{y}$, i.e., $||\bm{y}-\hat{\bm{y}}_k||^2= \operatorname*{min}\limits_{ k'\in \mathbb{K}}\{ \mathfrak{D}_{k'}  \}$, where $\mathfrak{D}_{k'} $ is the Euclidean distance between the actual received signal $\bm{y}$ and the corresponding constructed signal $\hat{\bm{y}}_{k'}$ with the channel $\bm{H}_{k'}$. 
Inspired by the above observations, 
the sender detection strategy can be interpreted form the perspective of the  generalized likelihood ratio test (GLRT) \cite{Kay1993Fundamental}, written as 

\begin{small}
\begin{equation}
\begin{split}
& P(\bm{Y};\mathcal{H}_1, \bm{\hat{y}}_1)=\frac{\mathrm{exp}\{-\frac{1}{2\sigma^2}(\bm{y}-\bm{\hat{y}}_1)^H (\bm{y}-\bm{\hat{y}}_1)    \}}{\sigma \sqrt{2\pi^{N_r}}},\\
&\vdots\\
& P(\bm{Y};\mathcal{H}_K, \bm{\hat{y}}_K)=\frac{\mathrm{exp}\{-\frac{1}{2\sigma^2}(\bm{y}-\bm{\hat{y}}_K)^H (\bm{y}-\bm{\hat{y}}_K)    \}}{\sigma \sqrt{2\pi^{N_r}}},
\label{eq:GLRT1}
\end{split}
\end{equation}
\end{small}%
where the hypothesis with the highest probability (the maximal likelihood) will be considered as the real sender.
Finally, we have the MLE-based sender detection strategy as 

\begin{small}
\begin{equation}
\begin{split}
 \mathcal{D}_{MLE}^{\ast}= \operatorname*{min}\limits_{k \in \mathbb{K}} \{ ||( \bm{I}_{N_r}-\bm{H}_1 \bm{H}_1^{\dag})\bm{y}||^2,..., ||( \bm{I}_{N_r}-\bm{H}_K \bm{H}_K^{\dag})\bm{y}||^2 \},
\label{eq:MLE detection 2}
\end{split}
\end{equation}
\end{small}%
where $\bm{I}_{N_r}-\bm{H}_k \bm{H}_k^{\dag} $ denotes the equivalent detector. Note that $\bm{H}_k \bm{H}_k^{\dag} \neq \bm{I}_{N_r}$ in strong receiver case with $N_r > N_t$. 



\subsection{The Case of a Strong Sender ($N_r\leq N_t$)}

In the case of a strong sender, the detection DoFs at the receiver are reduced. The multiplication of matrices $\bm{H}_k^H\bm{H}_k$ is rank-insufficient and thus the detector in \eqref{eq:MLE detection 2} becomes infeasible. 
A possible solution is to employ the well-known minimum mean square error (MMSE) estimator to estimate $\hat{\bm{x}}_k$, where the  sender detector becomes $ ||\big( \bm{I}_{N_r}- \bm{H}_k (\bm{H}_k^H \bm{H}_k+\frac{\sigma^2 N_t}{p}\bm{I}_{N_t} )^{-1}\bm{H}_k^H \big)  \bm{y}||$ with $p$ denoting the transmission power  at the sender.
Since the term $\frac{\sigma^2 N_t}{p}\bm{I}_{N_t}$ adds regularization effect for the matrix inverse operation, it makes the detector still feasible for the  $N_r\leq N_t$ configuration.
Nevertheless,  for the  MMSE based detector, the receiver needs the knowledge of the instantaneous transmission power $p$ at the sender, which could be difficult in practice and also the sender can simply keep varying its power to deteriorate the receiver's detection performance.
Hence, a more practical detection strategy is required for the strong sender case.
In this section, we alternatively propose  a maximum norm (M-Norm) based detector, as detailed below.




Starting from the fact that the norm of $\bm{H}_k^H \bm{H}_k$ is more likely to be larger than the norm of $\bm{H}_{k'}^H \bm{H}_k$, $\forall k'\neq k, k' \in \mathbb{K}$, it is safe to conclude that with high probability it holds that 
$||\bm{H}_k^H   \bm{H}_k \bm{W}_k \bm{s}_k||^2 \geq ||\bm{H}_{k'}^H   \bm{H}_k \bm{W}_k \bm{s}_k||^2$. 
Since the term $\bm{H}_k \bm{W}_k \bm{s}_k$ is the received signal excluding noise, it is intuitive to  multiply the received signal $\bm{y}$ with different $\bm{H}_k^H$ and calculate the norm of  $\bm{H}_k^H\bm{y}$, $\forall k \in \mathbb{K}$. If the signal indeed comes from the channel $\bm{H}_k$, the resulting norm should be the largest among all the candidates. 
Finally for the strong sender case, we reach a so-called M-Norm based sender detector  as 

\begin{small}
\begin{equation}
\begin{split}
\mathcal{D}_{M-Norm}^{\ast}:\operatorname*{max}\limits_{k\in \mathbb{K}} \{ ||\bm{H}_1^H \bm{y}||^2,...,||\bm{H}_K^H \bm{y}||^2\}.
\label{eq:MLE detection}
\end{split}
\end{equation}
\end{small}%


\subsection{Complexity Analysis of the Sender Detection Schemes}

Now we calculate the complexities of the detectors.
For the MLE based detector, its complexity is dominated by generating the pseudo-inverse matrices of the different MIMO channels. A pseudo-inverse matrix can
be obtained by the  singular value decomposition (SVD) approach or Cholesky decomposition \cite{Arakawa2006Computational}, which have been shown to offer similar complexity results, and the complexity is calculated as 
$16N_r^2N_t+24N_rN_t^2+29N_t^3$. Afterwards, it reconstructs the estimated version of the received signal and calculates the Euclidean distance to the real received signal $\bm{y}$, whose complexity is $8N_rN_t+8N_r$. Hence, the overall complexity of the MLE based detector is computed as $K(16N_r^2N_t+24N_rN_t^2+29N_t^3+8N_rN_t+8N_r)$. On the other hand, the M-Norm based detector multiplies the received signal with the different  $\bm{H}_k^H$ and compares the norm of $\bm{H}_k^H\bm{y}$ in sequence. Its overall complexity is given as $K(8N_tN_r+8N_t)$, which is evidently lower than that of the MLE based detector.

\section{Anonymous Precoding Design}
In section III, we have presented two novel sender detectors  that analyze the received signal together with the characteristics of MIMO channel to unmask the sender. 
In this section, on the contrary we investigate anonymous precoder design at the sender end, which judiciously manipulates the pattern of the received signal to inhibit the receiver's detection.
Again, we investigate anonymous precoding designs for strong receiver and strong sender cases, respectively.

\subsection{Anonymous Precoder for a Strong Sender Case ($N_r \leq N_t$) }

Since  the aim of sender anonymity  is to guarantee  receive quality for communications and meanwhile to conceal the sender's identity, a reasonable anonymous precoder needs to strike a good trade-off between these two metrics.
Before we give problem formulation, we first present Proposition 4.1 for the anonymous precoding design.

\textit{\textbf{Proposition 4.1}}: Implementing sender's anonymity conflicts with the design of receive equalizer, and thus anonymity is achieved at the cost of reduced receive diversity. ~~~~~~~~~~~~~~~~~~~$\square$

Proposition 4.1 can be proved by a counter example. If the receive performance can be enhanced by a channel equalizer at the receiver, no  anonymity can be achieved as the equalizer is built on acknowledging the real sender's MIMO channel. On the other hand, if anonymity is maintained and the identity of the sender is concealed, the receiver fails to  know the exact channel that the signal comes from, further indicating that a correct equalizer would be impossible.
Since the receiver's equalizer design conflicts with the  anonymity requirement, Proposition 4.1 essentially indicates that we need to treat each receive antenna as an individual receiver and impose per-antenna SINR constraint for multiplexing streams. In the following, two anonymous precoders are proposed for the strong sender case, respectively. 


\subsubsection{Interference-suppression based anonymous (ISA) Precoder}

Without loss of generality, assume the $k$-th user as the sender $\mathbb{S}$ at uplink.
For ease of expression, we simply write  $\bm{s}$ as the intended symbol vector and $\bm{W}$ as the associated precoder matrix. 
Since at most $N_r$ streams can be multiplexed in the strong sender case, we have $\bm{s} \in \mathbb{C}^{N_r \times 1}$ and $\bm{W}\in \mathbb{C}^{N_t \times N_r}$.
As revealed by Proposition 4.1, each receive antenna is treated as an individual receiver, and thus we assume the $i$-th receive antenna's desired symbol as $s_i$ from the vector $\bm{s}$, $\forall i \in N_r$.  
Denote  $\bm{q}_{i}\in \mathbb{C}^{N_t \times 1}$ as the $i$-th column of a precoding matrix $\bm{W}$ (i.e.,
$\bm{W}=[\bm{q}_1,...,\bm{q}_{N_r}]$), which corresponds to the precoder vector for the  symbol $s_{i}$. Denote $\bm{h}_i \in \mathbb{C}^{1 \times N_t}$ as the channel between the $i$-th receive antenna and sender (i.e., $\bm{H}_k=[\bm{h}_1^T,..., \bm{h}_{N_r}^T]^T$).
To scramble the proposed M-Norm detector  in section III, the anonymous precoder should guarantee the norm of
$\bm{H}_k^H\bm{y}$ small enough to address the norm test. Since the exact value of the receive noise is not known by the sender, we can alternatively suppress the value of $|| \bm{H}_k^H\bm{H}_k \bm{W}||^2$, which has the same effect of manipulating the norm of $|| \bm{H}_k^H\bm{y}||^2$ and guarantees the real sender  hiding in the user set $\mathbb{K}$.   
Now, we are able to present problem formulation, where we aim to maximize the minimal per-antenna SINR threshold $\Gamma$ under the power budget and anonymity constraints, such as 

\begin{small}
\begin{equation}
\begin{split}
 & P1: \operatorname*{max} \limits_{\bm{W}=[\bm{q}_1,...,\bm{q}_{N_r}]} ~\Gamma,~~\mathrm{s.t.} ~(C1):  \frac{||\bm{h}_i\bm{q}_i  ||^2}{\sigma^2+\sum_{i'=1, i'\ne i}^{N_r}||\bm{h}_{i}\bm{q}_{i'}  ||^2   } \geq  \Gamma, \forall i \in N_r,\\
 &~~~~~~~~~~~~~~~~~~~~~~~~~~~~~~~~~(C2): \sum_{i=1}^{N_r} ||\bm{q}_i||^2 \leq p_{max},~(C3):|| \bm{H}_k^H \bm{H}_k [\bm{q}_1,...,\bm{q}_{N_r}] ||^2 \leq \epsilon,
 \label{eq:Recevied signal P1}
\end{split}
\end{equation}
\end{small}%
where (C1) denotes that the per-antenna SINR  should be higher than the lower-bound $\Gamma$, which is the objective to be optimized. It is also observed that each receive antenna is impaired by inter-antenna interference, which acts as multi-user interference in  multiple-input and single-output (MISO) systems.
Constraint (C2) guarantees the dissipated transmission power lower than a budget $p_{max}$. Constraint (C3) suppresses the norm to be lower than a threshold $\epsilon$ to scramble the sender detector at the receiver.

The optimization P1 is a NP-hard problem and belongs the class of  non-convex  second-order cone programming (SOCP), where the coupling of the objective $\Gamma$ and inter-antenna interference   makes the optimization  intractable.
However, it is straightforward to show that the inequality power constraint (C2) will be achieved with equality at the
optimum. Otherwise, if there is power left, we can simply increase the transmission power to further improve the value of $\Gamma$, thus contradicting optimality.
Hence, we begin with the dual power minimization problem as

\begin{small}
\begin{equation}
\begin{split}
 & P1(a): \operatorname*{min}~~ f_{\Gamma^{(j)}}([\bm{q}_1,...,\bm{q}_{N_r}]) \triangleq \sum_{i=1}^{N_r} ||\bm{q}_i||^2\\
 &\mathrm{s.t.} ~(C4):  \frac{||\bm{h}_i\bm{q}_i  ||^2}{\sigma^2+\sum_{i'=1, i'\ne i}^{N_r}||\bm{h}_{i}\bm{q}_{i'} ||^2   } \geq \Gamma^{(j)}, \forall i \in N_r,~(C5):|| \bm{H}_k^H \bm{H}_k [\bm{q}_1,...,\bm{q}_{N_r}] ||^2 \leq \epsilon, 
 \label{eq:Recevied signal 1 P1a}
\end{split}
\end{equation}
\end{small}%
where $\Gamma^{(j)}$ serves as the per-antenna minimum SINR requirement and superscript $j$ denotes the index of iteration as detailed later.  
Let $f^{\ast} _{\Gamma^{(j)}} $ represent the optimal value of P1(a) with minimum SINR requirement $\Gamma^{(j)}$.
In fact, solving P1 with  (C2) upper bounded by $f^{\ast}_{\Gamma^{(j)}}$ yields an
optimal objective value of $\Gamma^{(j)}$. Furthermore, the optimal objective values of problems P1(a) and P1 are strictly monotonic increasing.
Therefore, considering $\Gamma^{(j)}$ as a variable of optimization, the optimal solution of P1 can be obtained by alternatively solving P1(a) for a
given $\Gamma^{(j)}$ and searching over different $\Gamma^{(j)}$. 
Since P1(a) is still a non-convex SOCP problem,  we  define $\bm{Q}_i=\bm{q}_i\bm{q}_i^H\in\mathbb{C}^{ N_t \times N_t}, \forall i \in N_r$, and transform P1(a) into a semi-definite programming (SDP) as

\begin{small}
\begin{equation}
\begin{split}
& P1(b):~ \operatorname*{min} \sum_{i=1}^{N_r}\mathrm{Tr}(\bm{Q}_i)~~\mathrm{s.t.} ~(\tilde{C4}): 
 \mathrm{Tr}(\bm{h}_i\bm{Q}_i \bm{h}_i^H)-\Gamma^{(j)}(\sigma^2+  \sum_{i'=1, i'\ne i}^{N_r} \mathrm{Tr}(\bm{h}_{i}\bm{Q}_{i'} \bm{h}_{i}^H)) \geq 0, \forall i \in N_r,\\
 &~~~~(\tilde{C5}): \mathrm{Tr}( \bm{H}_k^H \bm{H}_k  (\sum_{i=1}^{N_r}\bm{Q}_i   ) \bm{H}_k \bm{H}_k^H    ) \leq \epsilon,~(C6):\bm{Q}_i \succeq \bm{0},\forall i \in N_r, (C7): \mathrm{Rank}(\bm{Q}_i)=1,\forall i \in N_r,
 \label{eq:Problem P1b}
\end{split}
\end{equation}
\end{small}%
where  ($\tilde{\mathrm{C}4}$) and ($\tilde{\mathrm{C}5}$) are  linear matrix inequalities (LMI)s transformed from (C4) and (C5).
  (C6) and (C7) are the SDR version of $\bm{Q}_i=\bm{q}_i\bm{q}_i^H$, $\forall i \in N_r$.
Neglecting the rank-one constraint (C7), the problem P1(b) is defined as a ``separable SDP'' (SSDP) problem \cite{Boyd2004Convex}, which can be readily solved by convex optimization solvers. 
Hence, the procedure starts with an initial value of $\Gamma^{(j)}$, and we solve P1(b) to obtain the $\bm{Q}_i^{\ast}$, $\forall i \in N_r$. If the consumed power, i.e., $\sum_{i=1}^{N_r}\mathrm{Tr}(\bm{Q}_i)$, is smaller than the budget $p_{max}$, we can increase the value of $\Gamma^{(j)}$, otherwise decrease the value of $\Gamma^{(j)}$. The iteration is operated until convergence, as summarized in Algorithm 1.

\begin{algorithm}
\begin{small}
\caption{The Equivalence between non-convex SOCP  P1 and convex SDP P1(b) }
\label{alg:Algorithm1}
\begin{algorithmic}[1]
\renewcommand{\algorithmicrequire}{ \textbf{Input:}} 
\renewcommand{\algorithmicensure}{ \textbf{Output:}} 
\REQUIRE MIMO channel $\bm{H}_k$,  power budget $p_{max}$, symbol vector $\bm{s}$, initial left bound $\Gamma_{l}$, right bound $\Gamma_{r}$, anonymity threshold $\epsilon$,
and tolerance $\tau$.
\STATE Initialize $\Gamma^{(j)}=(\Gamma_{l}+\Gamma_{r})/2$.
\WHILE{$|\Gamma_{r}-\Gamma_{l}| \geq \tau $} 
\STATE Solve P1(b) with $\Gamma^{(j)}$. Let $f^{\ast}_{\Gamma^{(j)}}=\sum_{i=1}^{N_r}\mathrm{Tr}(\bm{Q}_i)$. Calculate the power reward factor ${R}=p_{max}-f^{\ast}_{\Gamma^{(j)}}$.
\IF{$R\geq 0$}
 \STATE Update $\Gamma_{l}=\Gamma^{(j)}$.
\ELSE
 \STATE Update $\Gamma_{r}=\Gamma^{(j)}$.
\ENDIF
\STATE Update the iteration index $j = j + 1$; Update $\Gamma^{(j)}= \frac{\Gamma_{l}+\Gamma_{r}}{2}$.
\ENDWHILE
\ENSURE Optimal SDP matrices $\bm{Q}_i^{\ast}$, $\forall i \in N_r$.
\end{algorithmic}
\end{small}
\end{algorithm}

After performing Algorithm 1, a non-trivial question is whether the obtained optimal solution  $\bm{Q}_i^{\ast}$ is of rank 1. 
Apparently, if it is, then the SDR relaxation is tight and the optimal beamformer  $\bm{q}_i^{\ast}$ can be simply obtained from the principal eigen-vector of $\bm{Q}_i^{\ast}$. Regarding the rank of the optimal solution $\bm{Q}_i^{\ast}$, $\forall i \in N_r$, we then have the following Proposition 4.2.

\textit{\textbf{Proposition 4.2}}:  Under the condition of independently distributed MIMO channels, the optimal solution of P1(b) satisfies $\mathrm{Rank}(\bm{Q}_i)=1, \forall i \in N_r$, with probability one. ~~~~~~~~~~~~~~~~~~~~~~~~$\square$

Proof: Please refer to APPENDIX A. $~~~~~~~~~~~~~~~~~~~~~~~~~~~~~~~~~~~~~~~~~~~~~~~~~~~~~~~~~~~~~~~~\blacksquare$

It is interesting that when the channels happen to be not independently distributed (e.g., in the case of line-of-sight or channel correlation), the tightness of the SDRs  can still be guaranteed in P1(b) by applying the rank reduction results in \cite{Huang2018Rank}, as summarized in Proposition 4.3.

\textit{\textbf{Proposition 4.3}}: Consider a SSDP \cite{Huang2018Rank} such as

\begin{small}
\begin{equation}
\begin{split}
& \mathrm{(SSDP)}:~ \operatorname*{min}_{\bm{X}_1,...,\bm{X}_L} \sum_{l=1}^{L}\mathrm{Tr}(\bm{B}_l\bm{X}_l)\\
 &\mathrm{s.t.}~  \sum_{l=1}^{L} \mathrm{Tr}(\bm{A}_{ul}\bm{X}_l  ) \unrhd_u b_u, u=1,...,U, \mathrm{and} ~\bm{X}_l \succeq \bm{0},l=1,...,L, 
 \label{eq:SSDP problem}
\end{split}
\end{equation}
\end{small}%
where $\bm{B}_l$ and $\bm{A}_{ul}$, $\forall l \in L, \forall u \in U$, are Hermitian matrices (but not necessarily positive semi-definite). $b_u \in \mathbb{R}$ and  $\unrhd_u \in \{ \leq, \geq,=   \}$  $\forall u \in U$.
Suppose that the SSDP is feasible and bounded, and the optimal value  is attained. There always exists an optimal solution $(\bm{X}_1^{\ast},...,\bm{X}_L^{\ast})$
such that $\sum_{l=1}^{L}\mathrm{Rank}^2(\bm{X}_l^{\ast}) \leq U$ \cite{Huang2018Rank}. ~~~~~~~~~~~~~~~~~~~~~~~~~~~~~~~~~~~~~~~~~~~~~~~~~~~~~~~~~~~~~~~~~~~~~~~$\square$

By applying this result in our context, it can be verified that $\sum_{1}^{N_r}\mathrm{Rank}^2(\bm{Q}_i^{\ast}) \leq N_r +1$. Also, it is evident from the per-antenna SINR constraint that  $\mathrm{Rank}(\bm{Q}_i^{\ast})\neq 0$, denoting that $\mathrm{Rank}(\bm{Q}_i^{\ast})\geq 1$, $\forall i \in N_r$. Hence, it is safe to include that there still exists a rank-1 solution such as $\mathrm{Rank}(\bm{Q}_i^{\ast})= 1$, $\forall i \in N_r$, which makes the SDRs of the Algorithm 1 still tight.
That is, if the obtained optimal result $\bm{Q}_i^{\ast}$ happens to have a high rank, the rank-reduction techniques in \cite{Huang2018Rank} can be applied  to obtain  rank-one solutions. 

Now the tightness of the SDRs has been confirmed by Propositions 4.2 and 4.3. Nevertheless, while the receive SINR and sender's anonymity can always be guaranteed, the received signal propagating through the equivalent channel $\bm{H}_k\bm{W}$ may have phase ambiguity, which impairs the de-modulation at the receiver. 
A conventional method is to adopt receive phase equalization to align the phase of the received signal with the desired symbol. However, since the sender's identity is concealed by the anonymous precoder and the receiver may not be able to declare  a correct channel, the conventional receive phase equalization is disabled in anonymous communications.
To this end, we further propose Proposition 4.4 for a novel transmit phase equalization.

\textit{\textbf{Proposition 4.4}}: 
With the optimal precoder $\bm{q}^{\ast}_i$ for the intended symbol $s_i$,
the desired signal at the $i$-th receive antenna is calculated as $\bm{h}_i\bm{q}^{\ast}_i s_i$, which should have the same phase to that of the desired symbol $s_i$ for de-modulation purpose.
Write $\bm{h}_i\bm{q}^{\ast}_i=|\bm{h}_i\bm{q}^{\ast}_i|e^{j\varphi_i}$, where $\varphi_i$ denotes the angle of the complex number $\bm{h}_i\bm{q}^{\ast}_i$.
Thus, the transmit phase equalization  is given as $\bm{q}^{\ast}_i=\bm{q}^{\ast}_ie^{-j\varphi_i}$, which makes the desired signal have exactly same phase to the desired symbol $s_i$ to avoid phase ambiguity without violating anonymity and  per-antenna SINR performance. ~~~~~~$\square$

Proof: Recalling (C4), the power of the desired signal remains unchanged after the equalization such as 
$||\bm{h}_i\bm{q}_i^{\ast} e^{-j\varphi_i}  ||^2=||\bm{h}_i\bm{q}_i^{\ast}  ||^2$. 
Also, based on the trigonometry property of norm operation, the power of the overall inter-antenna interference after equalization $||\sum_{i'=1, i'\ne i}^{N_r} \bm{h}_{i}\bm{q}_{i'}^{\ast} e^{-j\varphi_{i'}} ||^2$ is upper bounded by  $\sum_{i'=1, i'\ne i}^{N_r}|| \bm{h}_{i}\bm{q}_{i'}^{\ast} e^{-j\varphi_{i'}}    ||^2=\sum_{i'=1, i'\ne i}^{N_r}|| \bm{h}_{i}\bm{q}_{i'}^{\ast}  ||^2 $, denoting the obtained optimal per-antenna SINR remained unchanged. On the other hand, the sender's anonymity is also maintained after transmit phase equalization, as phase rotation of $\bm{q}_{i}$ has no impact on the trace of $\bm{Q}_{i}$, $\forall i \in N_r$. $~~~~~~~~~~~~~~~~~~~~~~~~~~~~~~~~~~~~~~~~~~~~~~~~~~~~~~~~~~~~~~~~~~~~~~~~~~~~~~~~~~~~~~~~~~\blacksquare$

Now we are able to devise the whole ISA precoder, as summarized in Algorithm 2. We first run Algorithm 1 to obtain the optimal  matrix $\bm{Q}_i^{\ast}$, and $\bm{q}_i^{\ast}$ is immediately obtained with $\mathrm{Rank}(\bm{Q}_i^{\ast})=1$, otherwise matrix reduction is conducted based on Propositions 4.2 and 4.3, $\forall i \in N_r$.
Afterwards, transmit phase equalization is applied  for removing the receiver's phase ambiguity without loss of the optimality of the SINR  and anonymity performance.

\begin{algorithm}
\begin{small}
\caption{The Overall ISA Precoder Design}
\label{alg:Algorithm2}
\begin{algorithmic}[1]
\renewcommand{\algorithmicrequire}{ \textbf{Input:}} 
\renewcommand{\algorithmicensure}{ \textbf{Output:}} 
\REQUIRE MIMO channel $\bm{H}_k$ and symbol vector $\bm{s}$.
\STATE Perform Algorithm 1 to obtained the SDR  matrices $\bm{Q}_i^{\ast}$, $\forall i \in N_r$.
\FOR{$i=1:N_r$}
\STATE Decompose $\bm{Q}_i^{\ast}$ to obtain the $\bm{q}_i^{\ast}$ if Rank$(\bm{Q}_i^{\ast})=1$; 
\STATE Otherwise do rank reduction for $\bm{Q}_i^{\ast}$ and then decompose $\bm{Q}_i^{\ast}$.
\ENDFOR
\STATE Do transmit phase equalization to avoid phase ambiguity.
\ENSURE Optimal precoding design $[\bm{q}_1^{\ast},...,\bm{q}_{N_r}^{\ast}]$.
\end{algorithmic}
\end{small}
\end{algorithm}

\subsubsection{Constructive-Interference based Anonymous (CIA) Precoding}

In part 1), we have proposed a SDR based anonymous precoding design, where the inter-antenna interference is strictly suppressed to guarantee the per-antenna SINR constraint. 
That is, the inter-antenna interference is treated as a harmful element, and any interference adds perturbation to the received signal. 
Following this principle, one needs to perform transmit phase equalization to constrain the per-antenna's symbol within a region around the nominal point in the modulated signal constellation, as illustrated in Fig. 2(a).
Nevertheless, since the transmitted symbols are known by the sender, it is judicious to jointly utilize the spatial correlation among the channels and the symbols
to be transmitted, based on the concept of constructive interference (CI) \cite{Christos2013Known}.
That is, the inter-antenna interference has potential to be utilized as a desired element to push the per-antenna desired signals away from the detection thresholds of the signal constellation, where the increased distance to the detection threshold of demodulation  benefits the per-antenna receiving performance. Let us start by demonstrating the concept of CI in the following Lemma 4.1, and  then we elaborate  CI for addressing anonymous precoding design. 
For notation simplicity we assume PSK modulation, nevertheless the following is applicable to multi-level modulations \cite{Christos2013Known}.

\begin{figure}
	\centering
	\includegraphics[width=3.4 in]{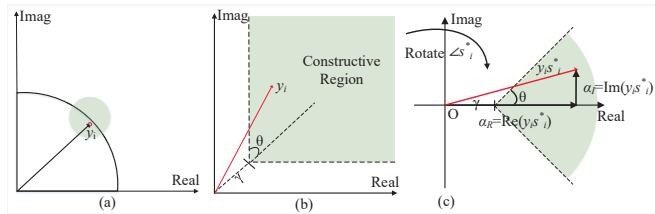}
    \caption{ The geometrical interpretation of CI precoding, where the intended symbol is $\frac{1 + i}{\sqrt{2}}$ with QPSK modulation for illustration. By CI design shown in Fig. 1(b), the received signal $y_i$ can be pushed into a constructive region (green area), rather than being strictly located in the proximity region around the constellation point. To guarantee the constructive effect of the interference, geometric interpretation can be exploited as shown in Fig. 1(c).}
    \label{fig:bpsk8psk} 
\end{figure}

\textit{\textbf{Lemma 4.1}}: Without loss of generality, write the intended symbol of the $i$-th receive antenna as $s_i=de^{j\phi_i}$ by M-PSK modulation, which can be further expressed as a rotated version of another symbol, such that $s_i=s_{i'}e^{j(\phi_i-\phi_{i'})}$. Hence, the received signal of the $i$-th receive antenna is written as 

\begin{small}
\begin{equation}
\begin{split}
&y_i=\bm{h}_i [\bm{q}_1,...,\bm{q}_{N_r}] \bm{s}+\bm{n}_i=\bm{h}_i \sum_{i'=1}^{N_r} \bm{q}_{i'} s_{i}e^{j(\phi_{i'}-\phi_{i})}+\bm{n}_i.
\label{eq:Recevied signal 1 ci}
\end{split}
\end{equation}
\end{small}%
Taking $s_1$ as a reference symbol, \eqref{eq:Recevied signal 1 ci} can be re-expressed as

\begin{small}
\begin{equation}
\begin{split}
&y_i=\bm{h}_ie^{j(\phi_{1}-\phi_i)} \sum_{i=1}^{N_r} (\bm{q}_{i'} e^{j(\phi_{i'}-\phi_1)})s_{i}+\bm{n}_i.
\label{eq:Recevied signal 2 ci}
\end{split}
\end{equation}
\end{small}%

Note that the reference symbol can be arbitrary. The reformulation in \eqref{eq:Recevied signal 2 ci} indicates that by exploiting the correlation among the channels and symbols rather than treating the input  as a Gaussian signal, the original inter-antenna interference channel reduces to a virtual multicast channel with common messages $s_i$ to all  receive antennas \cite{Wei2020Multi} \cite{Masouros2015Exploiting}. ~~~~~~~~~~~~~~~~~~~~~~~~~~~~$\square$


As suggested by Lemma 4.1,  inter-antenna interference can be utilized as a constructive element to benefit system performance, achieved by exploiting geometrical interpretation shown in Fig. 2.  Explicitly,  we first rotate the signal $y_i$ by the angle $\angle s^{\ast}_{i}$, and then the rotated signal can be mapped onto real axis $\alpha_{I}=\Im \{y_i s^{\ast}_i\}$ and imaginary axis $\alpha_{R} =\Re \{y_i s^{\ast}_i\}$, respectively. As can be seen, the received signal falls into a constructive region (in Fig. 1 (b)) if and only if the trigonometry $|\alpha_{I}| \leq (\alpha_{R}-\gamma) \mathrm{tan}\theta$ (in Fig. 1(c)) holds, where $\theta=\frac{\pi}{M} $ and $M$ represents constellation size.  In particular, $\gamma$  physically represents the Euclidean distance in the signal constellation between the constructive region and the decision thresholds, which also relates to SINR performance of the received signal, as depicted in Fig. 1(c). 
The above discussion can be extended into any order M-PSK and multi-level modulations.
For brevity we refer the readers to \cite{Christos2013Known} for details.
Hence, the inter-antenna interference can be  made constructive when the following inequality is satisfied.

\begin{small}
\begin{equation}
\begin{split}
& |\Im\{\bm{h}_i [\bm{q}_1,...,\bm{q}_{N_r}] \bm{s} s^{\ast}_i \}|\leq (\Re \{\bm{h}_i [\bm{q}_1,...,\bm{q}_{N_r}] \bm{s}s^{\ast}_i\} )-\gamma)\mathrm{tan}\theta,  \forall i \in N_r,
 \label{eq:CI effect 1}
\end{split}
\end{equation}
\end{small}%
which guarantees that the inter-antenna interference acts as a beneficial element to push the per-antenna received signal into constructive regions.
Nevertheless, when implementing CI with sender's anonymity, it is essential to impose additional anonymous constraint to manipulate the pattern of the received signal. Since the receiver adopts the M-Norm detector to unmask the sender, the following constraint is imposed to hide the sender  in the user set.

\begin{small}
\begin{equation}
\begin{split}
|| \bm{H}_k^H \bm{H}_k[[\bm{q}_1,...,\bm{q}_{N_r}]] \bm{s}||^2 \leq \zeta.
 \label{eq:anonymity constraint 1}
\end{split}
\end{equation}
\end{small}%
where $ \zeta$ serves as an anonymity-related threshold.
Now we are able to present the problem formulation for CI-based anonymous precoder. We target to maximize the value of $\gamma$, subject to multiple constraints. As discussed,  maximizing $\gamma$ equivalently optimizes the per-antenna receive performance, given as

\begin{small}
\begin{equation}
\begin{split}
 & P2: \operatorname*{max} \limits_{[\bm{q}_1,...,\bm{q}_{N_r}]} ~\gamma,~~\mathrm{s.t.}~ (C8): ||[\bm{q}_1,...,\bm{q}_{N_r}] \bm{s} ||^2\leq  p_{max},~(C9):|| \bm{H}_k^H \bm{H}_k[[\bm{q}_1,...,\bm{q}_{N_r}]] \bm{s}||^2 \leq \zeta,\\
 &~~~~~~~~~~~~~~~~~~~~~~~~~~~~~(C10): |\Im\{\bm{h}_i [\bm{q}_1,...,\bm{q}_{N_r}] \bm{s} s^{\ast}_i \}|\leq (\Re \{\bm{h}_i [\bm{q}_1,...,\bm{q}_{N_r}] \bm{s}s^{\ast}_i\} )-\gamma)\mathrm{tan}\theta,  \forall i \in N_r.
 \label{eq:Recevied signal P2}
\end{split}
\end{equation}
\end{small}%

Evidently, the standard convex optimization P2 can be solved directly, and the whole algorithm is summarized in Algorithm 3. More importantly, the  transmit phase equalization is not required as the per-antenna received signal has been  designed to  exactly fall into the constructive regions of the constellation, as summarized in the Remark 4.1.

\textit{\textbf{Remark 4.1}}: Based on the CIA precoder, the  received signal of each antenna has been  directly located into  constructive regions of the constellation. Hence, the receiver can demodulate the received signal directly, according to the amplitude and phase of the received signal. As a result, the proposed CIA precoder removes the need for receive or transmit equalization, while utilizing inter-antenna interference as a beneficial element without loss of anonymity. $~~~~~~~~~~~~~~~~~~~\square$

\subsection{Anonymous Precoder for a Strong Receiver ($N_r > N_t$) }

In this subsection, we further investigate the anonymous precoding for a strong receiver case. 
With the configuration of $N_r > N_t$, the SDR formulation is not feasible due to the insufficient transmit DoFs. Nevertheless, by the CI-based precoder,  more streams can still be multiplexed than the number of transmit antennas \cite{Masouros2015Exploiting}, and hence in the strong receiver case all the $N_r$ antennas can be efficiently utilized. 
On the other hand, as mentioned in Section III, in the strong receiver case the receiver employs the MLE sender detector in \eqref{eq:MLE detection 2} to unmask the sender, which considers the user with the minimum value of $||(\bm{I}_{N_r}-\bm{H}_k (\bm{H}_k^H\bm{H}_k)^{-1}\bm{H}_1^H)\bm{y}||^2$ as the real sender. Similar to the anonymous strategy applied in section IV-A, one may design the precoder to manipulate the norm  higher than a threshold, i.e., $||(\bm{I}_{N_r}-\bm{H}_
k(\bm{H}_k^H\bm{H}_k)^{-1}\bm{H}_k^H)\bm{y}||^2 \geq \zeta$.
However, this anonymous constraint confines a non-convex set. More importantly, with the unknown deterministic value of noise, the above constraint reduces to an alternative constraint $||(\bm{I}_{N_r}-\bm{H}_k (\bm{H}_k^H\bm{H}_k)^{-1}\bm{H}_k^H) (\bm{H}_k[\bm{q}_1,...,\bm{q}_{N_r}]\bm{s}) ||^2 \geq \zeta$, where the left hand boils down to 0 and the constraint makes no sense. 

In fact, since the receiver calculates the 
norm in \eqref{eq:MLE detection 2} in sequence and considers the one with the minimum value as the sender, we can artificially create a alias sender $k'$, and confines the following inequality as



\begin{small}
\begin{equation}
\begin{split}
&|| (\bm{H}_{k'}\bm{H}_{k'}^{\dag} -\bm{H}_k\bm{H}_k^{\dag})   \bm{H}_k [\bm{q}_1,...,\bm{q}_{N_r}] \bm{s}||^2\leq \delta, \forall k'\neq k,  k' \in \mathbb{K},
 \label{eq:anonymous constraint 14}
\end{split}
\end{equation}
\end{small}%
which physically denotes that the $k-$th and $k'$-th users are equally suspicious to the receiver by setting a small valued threshold $\delta$. However, imposing $K-1$ constraints in \eqref{eq:anonymous constraint 14} significantly reduces the DoFs of precoder design and thus may result in poor per-antenna SINR performance.  
To make a good trade-off between the per-antenna SINR and anonymity, the sender can randomly select a $k'$ from $\mathbb{K}$ as the alias sender. As a result, there will be only 1 constraint in \eqref{eq:anonymous constraint 14} without significantly degrading DoFs of precoder design. Also, the receiver still fails to declare the correct sender, as the real sender $k$ and the alias $k'$ are equally suspicious.


Similar to P2, while we maximize the effect of $\gamma$ to exploit the beneficial effect of inter-antenna interference, anonymity constraint is also imposed  against the MLE based sender detector.

\begin{small}
\begin{equation}
\begin{split}
 & P3: \operatorname*{max} \limits_{[\bm{q}_1,...,\bm{q}_{N_r}]} ~\gamma,~~\mathrm{s.t.}~ (C11): ||[\bm{q}_1,...,\bm{q}_{N_r}]\bm{s} ||^2\leq  p_{max},\\
 &~~~~~(C12): |\Im\{\bm{h}_i [\bm{q}_1,...,\bm{q}_{N_r}] \bm{s} s^{\ast}_i \}|\leq (\Re \{\bm{h}_i [\bm{q}_1,...,\bm{q}_{N_r}] \bm{s}s^{\ast}_i\} )-\gamma)\mathrm{tan}\theta,  \forall i \in N_r,\\
 &~~~~~(C13): || (\bm{H}_{k'}\bm{H}_{k'}^{\dag} -\bm{H}_k\bm{H}_k^{\dag}) \bm{H}_k [\bm{q}_1,...,\bm{q}_{N_r}] \bm{s}||^2\leq \delta,  k' \in \mathbb{K},
 \label{eq:Recevied signal P3}
\end{split}
\end{equation}
\end{small}%
which is a standard convex optimization and the whole algorithm is also included in Algorithm 3 for simplicity.

\begin{algorithm}
\begin{small}
\caption{The CIA Precoder Design}
\label{alg:Algorithm3}
\begin{algorithmic}[1]
\renewcommand{\algorithmicrequire}{ \textbf{Input:}} 
\renewcommand{\algorithmicensure}{ \textbf{Output:}} 
\REQUIRE MIMO channel $\bm{H}_k$,  power budget $p_{max}$, symbol vector $\bm{s}$.
\STATE Solve the standard convex optimization P2 (strong sender case).
\STATE Or solve the standard convex optimization P3 with a random alias sender (strong receiver case).
\ENSURE Optimal precoding design $[\bm{q}_1^{\ast},...,\bm{q}_{N_r}^{\ast}]$.
\end{algorithmic}
\end{small}
\end{algorithm}

\subsection{Complexity Analysis for the Anonymous Precoders}


In this subsection we investigate the complexities of the proposed precoders.
We first consider the strong sender case \footnote{For convex formulations that involve linear matrix inequality  (LMI) and SOC constraints, their complexities can be evaluated as $\mathrm{ln}(\frac{1}{\tau}) \sqrt{c_{b}}(c_{form}+c_{fact})$ \cite{Wang2014Outage}. 
Specifically, $\mathrm{ln}(\frac{1}{\tau})$ relates to the accuracy setup. $\sqrt{c_{b}}$ represents the barrier parameter measuring the geometric complexity of the conic constraints. $c_{form}$ and $c_{fact}$ represent the complexities cost on forming and factorization of $n\times n$ matrix of the linear system. We refer readers to \cite{Wang2014Outage} for details.}. For the ISA precoder, it first iteratively solves P1(b) to obtain the optimal SDR matrices $\bm{Q}_i$, $\forall i \in N_r$. Since P1(b) is subject to $N_r$ LMI constraints (trace) in ($\tilde{C4}$) with size 1,  1 LMI constraint (trace) in  ($\tilde{C5}$) with size 1, $N_r$ LMI constraints in (C6) with size $N_t$ (and (C7) is removed by SDR operation), the 
 complexity for iteratively optimizing P1(b) is given as 
$l_{i}\sqrt{N_r+1+N_rN_t} \mathrm{ln}(\frac{1}{\tau}) \big( n_1(N_r+1+N_rN_t^3)+n_1^2(N_r+1+N_rN_t^2) +n_1^3 \big)$, where $l_{i}$ denotes the number of iterations for convergence and will be further demonstrated in simulations. $\tau$ represents the tolerance of accuracy. 
Afterwards, eigenvalue decomposition for $\bm{Q}_i$ is computed for obtaining $\bm{q}_i$ with complexity $23N_t^3$, followed by transmit phase equalization with complexity $8N_t$. Hence, the overall complexity of the ISA anonymous precoder is given as $l_i\sqrt{N_r+1+N_rN_t} \mathrm{ln}(\frac{1}{\tau}) \big( n_1(Nr+1+N_rN_t^3)+n_1^2(N_r+1+N_rN_t^2) +n_1^3 \big)+N_r(23N_t^3+8N_t)$.
On the other hand, the CIA precoder (strong sender case) in (P2) is subject to 1 SOC constraint in (C8), $N_r$ linear constraints in (C9), and 1 SOC constraint in (C10). Hence, its overall complexity is given as
$\sqrt{4+N_r}\mathrm{ln} (\frac{1}{\tau}) \big( n_2N_r+n_2^2Nr+n_2(N_t^2+N_t^2) +n_2^3 \big)$.

Now we consider the complexity of the CIA precoder in the strong receiver case (P3). It is subject to 1 SOC constraint in (C11), $N_r$ linear constraints in (C12), and 1 SOC constraint in (C13). Hence, its overall complexity is given as 
$\sqrt{4+N_r}\mathrm{ln}(\frac{1}{\tau}) ( n_2N_r+n_2^2N_r+n_2(N_t^2+N_r^2) +n_2^3 )$. By comparing the complexities of the precoders, we have the following observation.  

\textit{\textbf{Remark 4.2}}:  Since the  per-antenna SINR constraint of the ISA precoder is imposed by the fractional-structured SOC constraints in (C1) of P1, it is further transformed into LMI constraints in P1(b). 
In comparison, by the CIA precoder, the per-antenna SINR constraint is imposed in the form of linear constraints ((C9) in P2 or (C12) in P3), which generally requires lower computational complexity over the LMI constraint in P1(b). Also, the CIA precoder directly locates the received signal at the receiver into constructive regions,
and hence the subsequent matrix decomposition and transmit phase equalization are not required. $~~~~~~~~~~~~~~~~~~~~~~~~~~~~~~~~~~~\square$

\begin{table*}
\centering
\centerline{TABLE \uppercase\expandafter{\romannumeral 1.}  Complexity analysis with accuracy factor $\tau$, where $n_1=\mathcal{O}(KN_t^2)$ and $n_2=\mathcal{O}(N_tN_r)$. }
\begin{tabular}{|c|c|c|c|} 
\hline
Sender & Strong sender& MLE  detector &$K(16N_r^2N_t+24N_rN_t^2+29N_t^3+8N_rN_t+8N_r)$ \\  
\cline{2-4}  
Detector& Strong  receiver& M-Norm detector &$K(8N_tN_r+8N_t)$\\
\cline{2-4}  
\hline
   &  & \multirow{2}*{ISA precoder}  & $l_i\sqrt{N_r+1+N_rN_t} \mathrm{ln}(\frac{1}{\tau}) \big( n_1(N_r+1+N_rN_t^3)+$\\ 
Anonymous  & Strong sender & &  $n_1^2(N_r+1+N_rN_t^2) +n_1^3 \big)+N_r(23N_t^3+8N_t)$\\
\cline{3-4}  
Precoder&  &CIA precoder&$\sqrt{4+N_r}\mathrm{ln} (\frac{1}{\tau}) \big( n_2 N_r+n^2N_r+2n_2N_t^2 +n_2^3 \big)$\\
\cline{2-4}  
& Strong receiver & CIA precoder &$\sqrt{4+N_r}\mathrm{ln}(\frac{1}{\tau}) ( n_2 N_r+n_2^2N_r+n_2(N_t^2+N_r^2) +n_2^3 )$\\
\cline{2-3}
\hline
\multirow{3}*{Benchmarks}&  &MMSE precoder \cite{Peel2005A} &$16N_r^2N_t+24N_rN_t^2+29N_t^3$\\  
\cline{3-4}  
&/ &SVD MIMO \cite{Tse2005Fundamentals} &$16N_r^2N_t+24N_rN_t^2+16N_rN_t^2+24N_t^3$ \\
\cline{3-4}  
& & CI precoder \cite{Masouros2015Exploiting}& $\sqrt{2+N_r}\mathrm{ln}(\frac{1}{\tau}) ( n_2 N_r+n_2^2N_r+ n_2N_t^2 +n_2^3 )$\\
\hline
\end{tabular}
\end{table*}


\section{Simulation Results}

We present the Monte-Carlo simulation results in this section. Without loss of generality, power budget is set to as $p_{max}=1$ Watt. QPSK is adopted as modulation scheme and the  symbol vector is
randomly generated. Assume that each block consists of 50 symbols.
There are $K=5$  senders, and the communication sender in each time slot (block) is randomly generated.  Rayleigh block-fading channel is adopted. 
The antenna configuration is set to as  $N_r=N_t=10$ in the strong sender case, while it is assumed that $N_r=10$ and $N_t=9$ in the strong receiver case. 
The energy detection threshold in \eqref{eq:Pf}
is set to as $\beta=10^{-2}$.
As revealed in section III, the receiver attempts to identify the real sender from the $K$ candidates, by employing the MLE/M-Norm sender detectors at the strong receiver/sender cases, respectively.
In addition, the following classic precoders are selected as benchmarks: 1) SVD precoder \cite{Tse2005Fundamentals}, where the receiver first detects the origin of the received signal and then calculates its receive equalizer based on the declared hypothesis. 2) MMSE  \cite{Peel2005A} and 3) CI precoder \cite{Masouros2015Exploiting}, where each receive antenna is treated as an individual receiver for multiplexing and hence no equalizer is required at the receiver.

\subsection{Strong Sender Case}

\begin{figure}
	\centering
	\includegraphics[width=3.2 in]{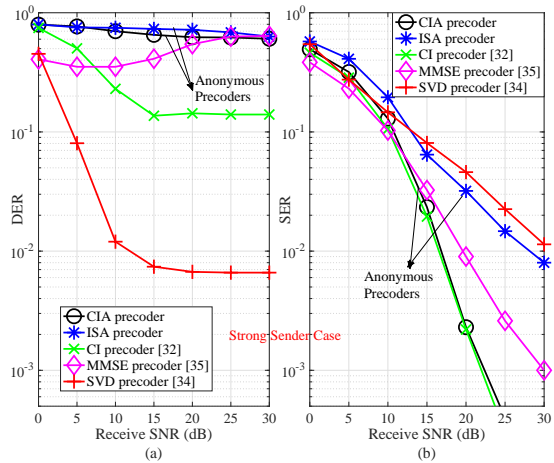}
    \caption{Strong sender case: the impact of receive SNR on the DER and SER by different precoders, where $N_t=N_r=10$, $\epsilon=20$ and $\zeta=8$.}
    \label{fig:DER_strongTX} 
\end{figure}

In Fig. \ref{fig:DER_strongTX}(a), the sender detection error rate (DER) performance of different precoders is demonstrated. It is observed that both the proposed anonymous ISA and CIA precoders achieve strong anonymity performance, where the receiver's DER performance is maintained at up to 0.8 even with high receive SNR. 
For the ISA precoder, its anonymity constraint is guaranteed by (C3) in P1, which is further transformed into a LMI constraint ($\tilde{C5}$). 
It can be seen that the anonymity is well guaranteed after the SDR operation of P1(b) and transmit phase equalization, confirming the 
analysis in Proposition 4.4. Also for the CIA precoder, the anonymity constraint (C10) in  P2 manipulates the pattern of the received signal to scramble the sender detection, which makes the receiver have a high probability of mis-clarifying the real sender.
In comparison, the SVD MIMO demonstrates the worst anonymity performance, where the receiver is able to unmask the correct sender with below $10^{-2}$ DER at 10 dB SNR. 
With the MMSE precoder, the receiver's DER demonstrates a U-shape when the receive SNR increases. It is because at low SNR regime, its detection performance is impaired by the receive noise. While at high SNR regime, the structure of the MMSE precoder approaches that of the ZF precoder such as $\bm{H}^H(\bm{H}\bm{H}^H)^{-1}$, and thus the received signal tends to be $\bm{y}=\bm{s}+\bm{n}$, where the sender's channel information is removed. As a result, the DER by  MMSE precoder is occasionally maintained at a high receive SNR regime. 
Also for the CI precoder, its target is to maximize the receive SINR performance without the consideration of sender's anonymity. In particular,  since it has been reported that the CI precoder is reduced to the ZF precoder when occasionally no interference can be exploited \cite{Li2018Interference}, a higher DER is achieved over the SVD precoder but is still less-anonymous to the proposed CIA precoder.
Last but not least, with the reduced detrimental impact of noise at higher SNR regimes,  the accuracy of the sender detector of the receiver is improved (except MMSE) and hence the receiver's detection becomes more accurate, resulting a decreased  DER performance.


In Fig. \ref{fig:DER_strongTX}(b), the symbol error rate (SER) performance under different precoders is demonstrated. 
Since the CI precoder is able to utilize the inter-antenna interference without anonymous constraints, the high DoFs at the sender side endorse the lowest SER performance among all the precoders \cite{Li2018Interference}. However, it can be seen that the proposed CIA precoder achieves a close SER performance to the CI precoder, and significantly outperforms the SVD and MMSE at moderate/high SNR regimes.
For the ISA precoder, although its DoF  of the precoder design is constrained by the anonymity constraint, it still demonstrates a close SER to the SVD precoder at 0-12 dB  SNR regimes, and outperforms the SVD precoder with above 12 dB SNR.
Hence, the two anonymous precoders indeed strike a good trade-off between  guaranteeing high communication quality and addressing sender's anonymity.

\begin{figure}
	\centering
	\includegraphics[width=3.2 in]{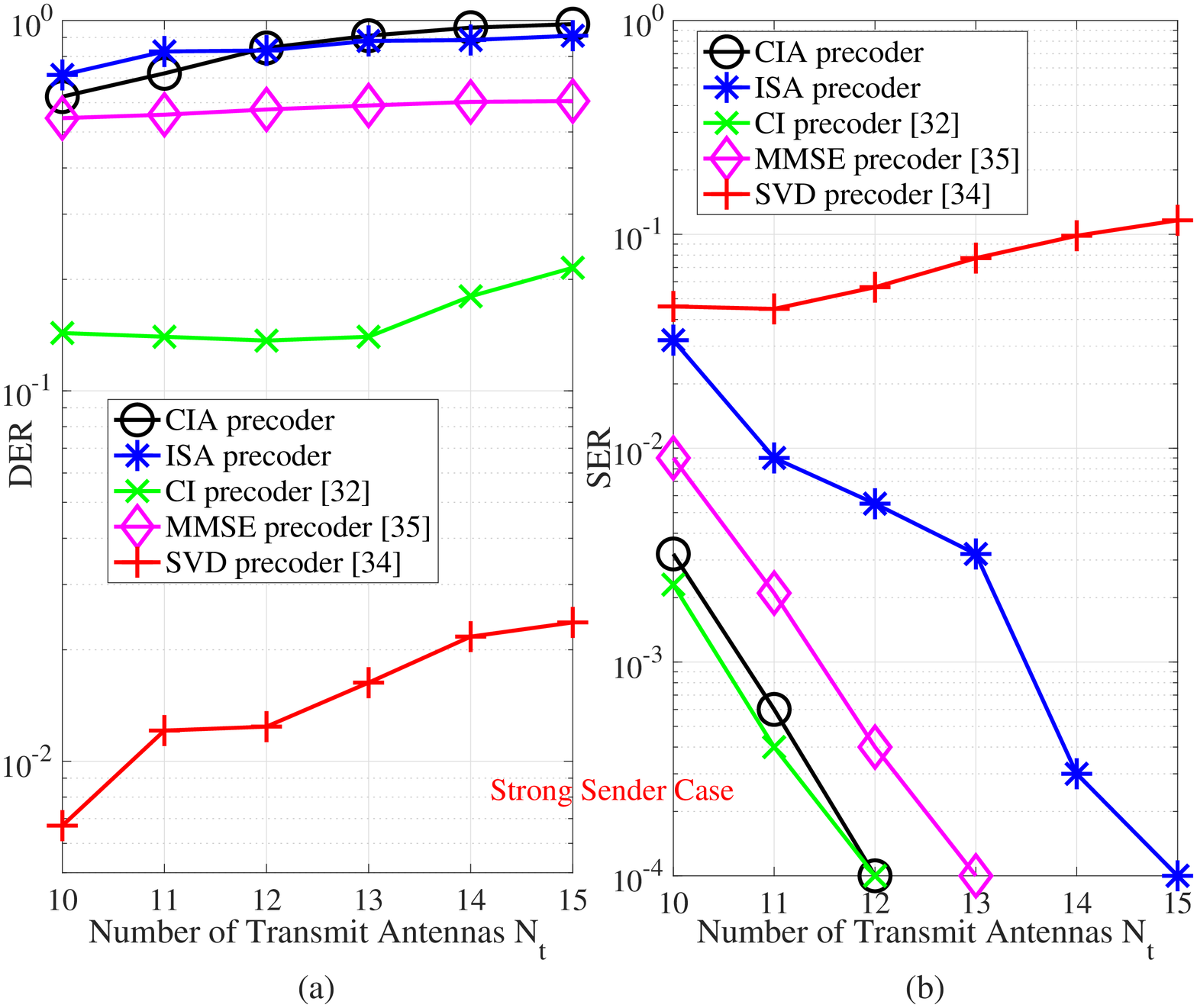}
    \caption{Strong sender case: the impact of different antenna configurations
    on the DER and SER performance by different precoders, where $N_r=10$,  $\epsilon=20$, $\zeta=8$, and SNR is fixed at 20 dB.}
    \label{fig:strongTX_antennas} 
\end{figure}

In Fig. \ref{fig:strongTX_antennas}, the DER and SER performances with different antenna configurations are demonstrated, where SNR is fixed at 20 dB. 
For the DER performance in Fig. \ref{fig:strongTX_antennas}(a), it is first observed that with more antennas, the DER of the anonymous precoders are improved. It is because with the increased dimension of a channel matrix, the impact of the anonymity  threshold $\epsilon$ in (C5) and $\zeta$ in (C10) becomes stricter, which leads to a more stringent anonymity requirement. Also, a similar trend can be observed by the benchmarks 
with distinct reasons. To be specific, with more transmit antennas, the spatial orthogonality of the MIMO channel between the sender and receiver is increased, and thus the structures of MMSE, SVD and CI precoders slightly tends to that of the ZF precoder. As a result, the DER performance of the benchmarks is increased with more transmit antennas, whereas the receiver is still  able to  declare the correct  sender with a high probability.
Second, with different antenna configurations, the proposed precoders always endorse a stricter anonymity compared to the benchmarks, where  the receiver is able to declare the correct sender with a DER lower than 0.03 by SVD precoder, with a DER lower than 0.3 by CI precoder, and with a DER lower than 0.6 by MMSE precoder.
For the SER performance Fig. \ref{fig:strongTX_antennas}(b), the CIA  precoder demonstrates a close performance to that of CI precoder, while the ISA precoder always outperforms the SVD precoder.
It is because with the increased transmit DoF, it is easier for the anonymous precoder to satisfy the anonymity constraint without much sacrificing the per-antenna SINR performance. 
In addition, it is worth noting that since the SVD's combiner is based on the sender detection performance, its SER first increases due to the high transmit DoF but begins to increase with $N_t\geq 12$ due to the improved DER performance.

\begin{figure}
	\centering
	\includegraphics[width=3.2 in]{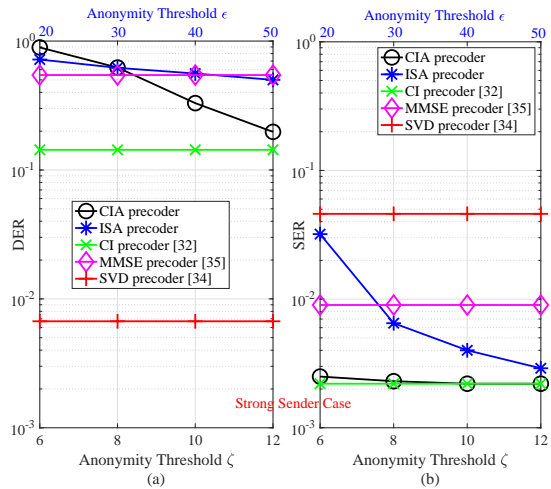}
    \caption{The impact of different values of the anonymity thresholds
    on the DER and SER performance, where $N_r=N_t=10$, and SNR is fixed at 20 dB.}
    \label{fig:different_threshold_TX} 
\end{figure}

Fig. \ref{fig:different_threshold_TX} demonstrates the effect of the anonymity thresholds on the DER and SER performance. It is observed for the proposed anonymous precoders, a smaller value of threshold endorses a stricter anonymity performance, while a loose anonymity threshold allows a better SER performance. It indicates that  a trade-off between the communication and anonymity can be  struck by setting different values for the thresholds, based on the stringency of the anonymity requirement. In particular, by selecting proper thresholds for the proposed ISA precoder (i,e., $\epsilon=30-45$) and CIA precoder (i.e.,  $\zeta=6-8$), they  outperform all the benchmarks in anonymity, and obtain lower SER over the MMSE and SVD precoders.

\begin{figure}
	\centering
	\includegraphics[width=3.0 in]{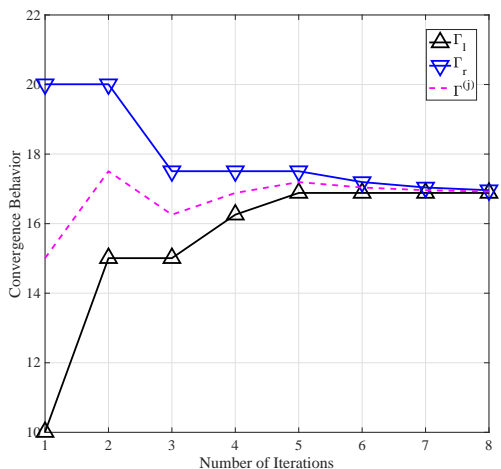}
    \caption{ The convergence behavior on finding $\Gamma^{(j)}$ by P1(b), where  tolerance factor $\tau=0.1$, $N_t=N_r=10$, and $\epsilon=20$.}
    \label{fig:convergence} 
\end{figure}

Fig. \ref{fig:convergence} shows the number of iterations by the ISA precoder for achieving convergence, with initial right bound $\Gamma_r=20$ and left bound $\Gamma_l=0$.
Since the bisection search requires at most $\mathrm{ln} (\frac{\Gamma_r-\Gamma_l}{\tau})$ iterations for convergence, it is seen that the algorithm converges to a stationary point with around 6-7 iterations,  confirming the low complexity of the ISA design.

\subsection{Strong Receiver Case}

\begin{figure}
	\centering
	\includegraphics[width=3.2 in]{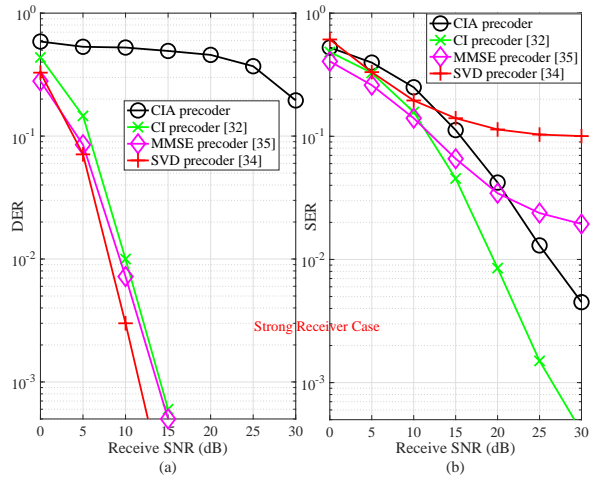}
    \caption{Strong receiver case: the impact of receive SNR on the DER and SER of the different precoders, where $N_t=9$, $N_r=10$, and $\delta=0.03$.}
    \label{fig:DER_strongRX} 
\end{figure}

Now we present the simulation results  in the strong receiver case. 
In Fig. \ref{fig:DER_strongRX}(a), the DER performance of different precoders is presented with various receive SNR. First, since the DoF of the receiver is improved with the strong receiver configuration, the MLE  detector can be employed, which helps the receiver obtain a more accurate detection performance compared to the strong sender case.
Importantly, it can be seen that the proposed CIA precoder still guarantees the DER above 0.5 at 0-15 dB SNR, above 0.4 at 15-25 SNR, and above 0.2 at 30 dB SNR regimes.
In comparisons, by the CI, MMSE and SVD  benchmarks, the receiver can correctly identify the real sender with $10^{-1}$-$10^{-2}$ DER at 5-10 dB SNR regimes, which further decreases to $10^{-3}$ with above 12 dB SNR regimes. 
In particular, with the strong receiver case $N_r>N_t$, the multiplication of the MIMO channel and MMSE precoder does not lead to an identity matrix such that $\bm{H}\bm{H}^H (\bm{H}\bm{H}^H+\frac{N_r\sigma^2}{p}\bm{I}_{N_r})^{-1} \neq \bm{I}_{N_r}$ due to the rank-sufficient property of $\bm{H}\bm{H}^H$. As a result, the channel information is not null-ed as that in the strong sender case, and thus the sender's anonymity is leaked to the receiver by the MMSE precoder.


In Fig. \ref{fig:DER_strongRX}(b), the SER performance of different precoders is presented with various receive SNR. 
It is shown that the proposed CIA  precoder outperforms the SVD precoder with above 11 dB receive SNR and the MMSE precoder with above 15 dB receive SNR, respectively. 
Also, although anonymity constraint limits the DoFs of the anonymous precoding design, the CIA precoder still provides a comparable SER performance to the CI precoder, and it demonstrates 5 dB SNR gain between the two precoders at $10^{-2}$ SER level.
Furthermore, it is worth mentioning the SVD precoder only supports $N_t$ streams in the strong receiver case, 
while the CIA precoder enables more data streams ($N_r$) than the number of the transmit antennas ($N_t$), confirming its applicability in both strong receiver/sender cases.

\begin{figure}
	\centering
	\includegraphics[width=3.2 in]{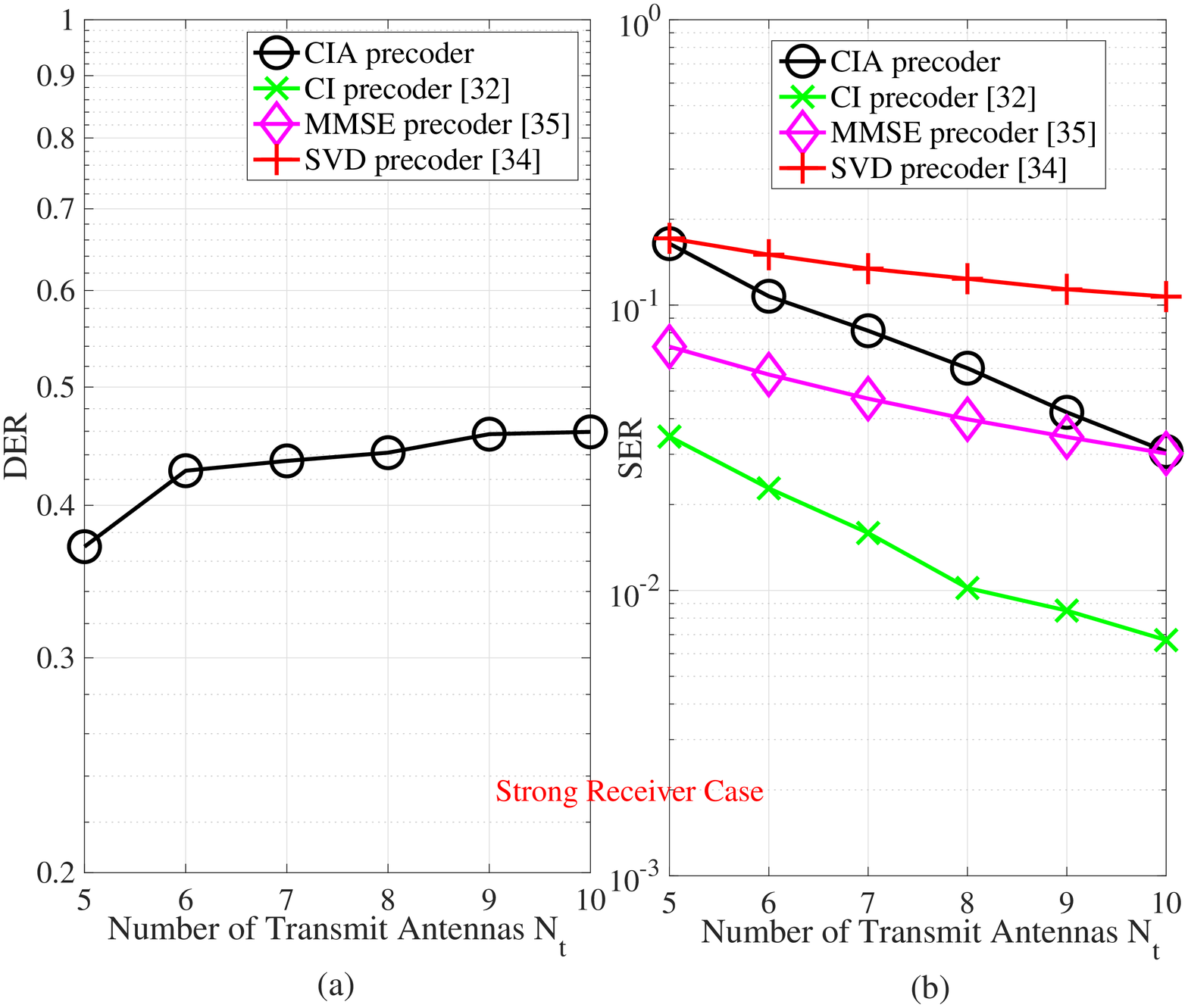}
    \caption{Strong receiver case: the impact of different antenna configurations
    on the DER and SER performance by different precoders, where $N_r=N_t+1$, $\delta=0.03$, and SNR is fixed at 20 dB.}
    \label{fig:strongRX_antennas} 
\end{figure}

\begin{figure}
	\centering
	\includegraphics[width=3.2 in]{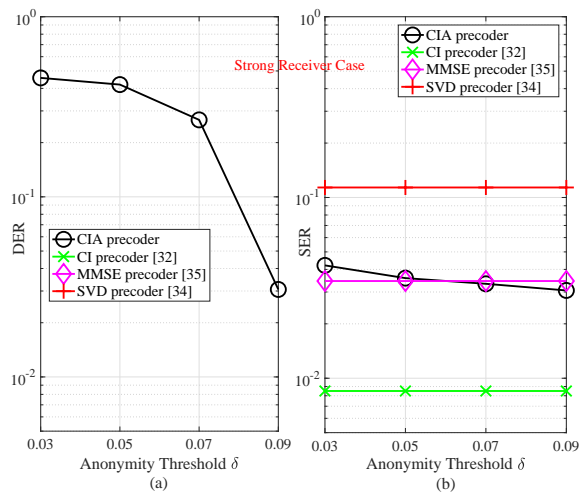}
    \caption{The impact of different values of the anonymity thresholds
    on the DER and SER performance, where $N_r=10$ and $N_t=9$. SNR is fixed at 20 dB.}
    \label{fig:different_threshold_RX} 
\end{figure}

In Fig. \ref{fig:strongRX_antennas}, the DER and SER performance with different antenna configurations are demonstrated in the strong receiver case, where SNR is fixed at 20 dB. 
In Fig. \ref{fig:strongRX_antennas}(a), the DER of the CI, MMSE and SVD precoder is reduced to 0, where the MLE based sender detector can perfectly clarify the real sender. As a comparison, the proposed anonymous precoder maintains the DER at a high level, which is around 0.4-0.5 with different numbers of the transmit antennas. 
On the other hand, Fig. \ref{fig:strongRX_antennas}(b) shows that the SER of the  proposed CIA precoder is always superior over the SVD precoder.  Also, with a higher DoF at the transmitter side, the SER of the anonymous precoder gradually approaches that of the MMSE and CI precoders, while the two benchmarks fail to address the anonymity as observed in  Fig. \ref{fig:strongRX_antennas}(a).

Fig. \ref{fig:different_threshold_RX} demonstrates the trade-off between the communication and anonymity performance with different anonymity thresholds in the strong receiver case.   
Since by the benchmarks, the receiver can perfectly declare the correct sender, their DER is virtually zero. 
In comparison with a proper threshold, i.e., $\delta=0.05$, the proposed CIA precoder maintains an above 0.4 DER performance, 
and also outperforms the  MMSE and SVD precoders in SER performance.

\section{CONCLUSIONS}

In this paper, we have proposed the concept of PHY anonymity, and revealed that by only analyzing PHY information, the receiver is able to unmask the sender's identity. With different antenna configurations, we have proposed two sender detection strategies for the receiver, one MLE  detector for the strong receiver case and one M-Norm detector for the strong sender case. Subsequently, we have investigated anonymous precoding design to guarantee the sender's anonymity while maximizing per-antenna SINR performance. Hence, we have further proposed an ISA precoder with tight SDR, assisted by a dedicated transmit phase equalizer for removing phase ambiguity, and a CIA precoder with the ability of utilizing inter-antenna interference as an useful source for improving SINR performance. Furthermore, the CIA precoder is also applicable to the strong receiver case, where more streams can be multiplexed than the number of transmit antennas without losing the sender's anonymity. 
Compared to the benchmarks,  simulation results have confirmed that the proposed anonymous precoders are able to mask the sender's identity, while simultaneously providing high per-antenna SINR for anonymous communications.

\begin{appendices}
	
\section{PROOF OF PROPOSITION 4.2}

The relaxed version of transformed problem P1(b) in \eqref{eq:Problem P1b} is
jointly convex with respect to the optimization variables and satisfies the Slater’s constraint qualification (without (C7)). 
Hence, strong duality holds and solving the dual problem is equivalent to solving the primal problem \cite{Sun2016Multi}. For obtaining the dual problem, we  write the Lagrangian function of \eqref{eq:Problem P1b} as

\begin{small}
\begin{equation}
\begin{split}
& \mathcal{L}=\sum_{i=1}^{N_r}\mathrm{Tr}(\bm{Q}_i)+ \mu (  \mathrm{Tr}(\bm{\Pi} \sum_{i=1}^{N_r} \bm{Q}_i)-\epsilon)
-\sum_{i=1}^{N_r} \bm{P}_i\bm{Q}_i + \sum_{i=1}^{N_r} \lambda_i \big( \Gamma^{(j)} \sigma^2+ \Gamma^{(j)}\sum_{i'\ne i, i'=1}^{N_r} \mathrm{Tr}(  \bm{G}_{i} \bm{Q}_{i'}    ) - \mathrm{Tr}( \bm{G}_i \bm{Q}_{i}   )     \big),
\label{eq:Lagrangian 1}
\end{split}
\end{equation}
\end{small}%
where $\bm{\Pi}=\bm{H}_k^H\bm{H}_k\bm{H}_k^H  \bm{H}_k$ and 
$\bm{G}_i=\bm{h}_i^H\bm{h}_i$ for brevity. $\mu$ and $\lambda_i$ are the Lagrange multipliers associated with constraints ($\tilde{C5}$) and ($\tilde{C4}$), respectively, while
matrix $\bm{P}_i \in \mathbb{C}^{N_t \times N_t}$ is the Lagrange multiplier matrix for the positive semi-definite constraint (C6). Hence, the dual problem for the optimization  in \eqref{eq:Problem P1b} is written as $\operatorname*{max}_{\mu \geq 0, \lambda_i \geq 0, \bm{P}_i \succeq \bm{0}} \operatorname*{min}_{\bm{Q}_i } 
\mathcal{L} (  \mu, \lambda_i, \bm{P}_i, \bm{Q}_i )$.

We reveal the structure of the optimal $\bm{Q}_i$ of (14) by studying the Karush-Kuhn-Tucker (KKT) conditions, which includes the dual constraints:
$\mu^{\ast} \geq 0, \lambda_i^{\ast} \geq 0, \bm{P}_i^{\ast} \succeq \bm{0}, \forall i \in N_r$; and complementary slackness: $\bm{P}_i^{\ast}\bm{Q}_i^{\ast} \succeq \bm{0}, \forall i \in N_r$;
and  the gradient of Lagrange function with respect to $\bm{Q}_i$ vanishing to 0: $\frac{\partial\mathcal {L}}{\partial \bm{Q}_i }|_{\bm{Q}_i^{\ast}}=0$: $\frac{\partial\mathcal {L}}{\partial \bm{Q}_i }|_{\bm{Q}_i^{\ast}}=\bm{I}_{N_t}
+ \Gamma^{(j)}\sum_{i'\ne i}^{N_r} \lambda_{i'}^{\ast} \bm{G}_{i'}+\mu^{\ast} \bm{\Pi}-\bm{P}_i- \lambda_i^{\ast}\bm{G}_i=0, \forall i \in N_r$, which further yields $\bm{P}_i^{\ast}=\bm{R}_i^{\ast}
- \lambda_i^{\ast}\bm{G}_i$, where $\bm{R}_i^{\ast}=\bm{I}_{N_t}
+ \Gamma^{(j)}\sum_{i'\ne i}^{N_r} \lambda_{i'}^{\ast} \bm{G}_{i'}+\mu^{\ast} \bm{\Pi}$. Indeed, it can be verified that in order to meet the per-antenna SINR constraints, it must hold that $\mathrm{rank}(\bm{Q}_i^{\ast} )\geq 1$ with $\bm{Q}_i^{\ast}\neq \bm{0}$. Hence, the complementary slackness $\bm{P}_i\bm{Q}_i = \bm{0}$ indicates

\begin{small}
\begin{equation}
\begin{split}
\mathrm{Rank}(\bm{P}_i^{\ast})\leq N_t -1.
\label{eq:rank 11}
\end{split}
\end{equation}
\end{small}%

If $\mathrm{Rank}(\bm{P}_i^{\ast}) = N_t-1$, then the optimal beamforming matrix $\bm{Q}_i^{\ast}$ must be a rank-one matrix. 
In order to further reveal the structure of $\bm{P}_i^{\ast}$, we first show by contradiction that $\bm{R}_i^{\ast}$ is a positive-definite matrix with probability one under the condition stated in the Proposition 4.2. 
For a given set of optimal dual variables, i.e., 
$\mu^{\ast}, \lambda_i ^{\ast}, \bm{P}_i^{\ast}$,
the dual problem  can be written as $\operatorname*{min}_{\bm{Q}_i } 
\mathcal{L} ( \bm{Q}_i, \mu^{\ast}, \lambda_i^{\ast}, \bm{P}_i^{\ast}  )$.
Suppose $\bm{R}_i^{\ast}$ is not positive-definite. In this case, we can choose $\bm{Q}_i=\beta \bm{r}_i \bm{r}_i^H$ as one of the optimal solution of the dual problem, where $\beta > 0$ is a scaling parameter and $\bm{r}_i$ is the eigenvector corresponding to a non-positive eigenvalue $\rho_i<0$ of  $\bm{R}_i^{\ast}$, i.e., $ \bm{R}_i^{\ast} \bm{r}_i=\rho_i\bm{r}_i$. Then, substituting $\bm{Q}_i=\beta \bm{r}_i \bm{r}_i^H$ and $ \bm{R}_i^{\ast} \bm{r}_i=\rho_i\bm{r}_i$ into the dual problem yields

\begin{small}
\begin{equation}
\begin{split}
\sum_{i=1}^{N_r}\mathrm{Tr}(\beta \bm{r}_i \bm{r}_i^H)-
\rho \sum_{i=1}^{N_r} \mathrm{Tr}( \bm{r}_i\bm{r}_i^H (\bm{P}_i^{\ast}+\lambda_i\bm{G}_i)  ), 
\label{eq:dual 33}
\end{split}
\end{equation}
\end{small}%
where the first term  is not positive. For the second term, since the channel vector $\bm{h}_i$ is  statistically independent, and based on  $\bm{P}_i^{\ast} \succeq \bm{0}$, we have the second term $\rho \sum_{i=1}^{N_r} \mathrm{Tr}( \bm{r}_i\bm{r}_i^H (\bm{P}_i^{\ast}+\lambda_i\bm{G}_i)  )$ is greater than 0. Setting $\rho \rightarrow \infty$, we have the term $-\rho \sum_{i=1}^{N_r} \mathrm{Tr}( \bm{r}_i\bm{r}_i^H (\bm{P}_i^{\ast}+\lambda_i\bm{G}_i)  ) \rightarrow -\infty $, where the dual optimal value becomes unbounded from
below. However, the optimal value of the primal problem \eqref{eq:Problem P1b} is non-negative. Thus, strong duality cannot hold which leads to a contradiction \cite{Sun2016Multi}. 
Therefore, $\bm{R}_i^{\ast}$ is a positive-definite matrix with probability one, i.e., $\mathrm{Rank}(\bm{R}_i^{\ast}) = N_t$. By applying $\bm{P}_i^{\ast}=\bm{R}_i^{\ast}
- \lambda_i^{\ast}\bm{G}_i$ and the sub-additivity  property of the rank operation, we have

\begin{small}
\begin{equation}
\begin{split}
\mathrm{Rank}(\bm{P}_i^{\ast})+\mathrm{Rank}(\lambda_i \bm{G}_i) \geq \mathrm{Rank}(\bm{P}_i^{\ast}+ \lambda_i \bm{G}_i) = \mathrm{Rank}(\bm{R}_i^{\ast})=N_t \Rightarrow \mathrm{Rank}(\bm{P}_i^{\ast})= N_t-1.
\label{eq:rank 23}
\end{split}
\end{equation}
\end{small}%

Based on \eqref{eq:rank 11} and \eqref{eq:rank 23}, we obtain that $\mathrm{Rank}(\bm{P}_i^{\ast})=N_t-1$. Thus, $\mathrm{Rank}(\bm{Q}_i^{\ast})=1$ holds with probability one.

\end{appendices}

\ifCLASSOPTIONcaptionsoff
  \newpage
\fi

\end{document}